\documentclass[a4paper,fleqn]{cas-dc}

\usepackage[numbers]{natbib}
\usepackage{multirow}

\newcommand{\cpgroup}{{CP~Group}}
\newcommand{\cpgroups}{{CP~Groups}}

\def\tsc#1{\csdef{#1}{\textsc{\lowercase{#1}}\xspace}}
\tsc{WGM}
\tsc{QE}
\tsc{EP}
\tsc{PMS}
\tsc{BEC}
\tsc{DE}

\begin{document}
\begin{sloppypar} 
\let\WriteBookmarks\relax
\def\floatpagepagefraction{1}
\def\textpagefraction{.001}
\shorttitle{Performance Scaling}
\shortauthors{S. Mandal et~al.}

\title [mode = title]{Improving the Scaling and Performance of Multiple Time Stepping based
Molecular Dynamics with Hybrid Density Functionals}                      

\author[1,2,3]{Sagarmoy Mandal}
\ead{mandal.sagarmoy@fau.de}


\address[1]{Department of Chemistry, Indian Institute of Technology Kanpur (IITK), 208016 Kanpur, India}

\address[2]{Interdisciplinary Center for Molecular Materials and Computer Chemistry
Center, Friedrich-Alexander-Universit\"at Erlangen-N\"urnberg (FAU),
N\a"gelsbachstr. 25, 91052 Erlangen, Germany}

\address[3]{Erlangen National High Performance Computing Center (NHR@FAU),
Friedrich-Alexander-Universität Erlangen-Nürnberg, Martensstr. 1, 91058
Erlangen, Germany}

\author[1]{Ritama Kar}

\ead{kritama@iitk.ac.in}


\author[2,3]{Tobias Klöffel}
\ead{tobias.kloeffel@fau.de}

\author[2,3]
{Bernd Meyer}[orcid=0000-0002-3481-8009]
\ead{bernd.meyer@fau.de}

\author%
[1]
{Nisanth N. Nair}[orcid=0000-0001-8650-8873]
\cormark[1]
\ead{nnair@iitk.ac.in}


\cortext[cor1]{Corresponding author}

\begin{abstract}
Density functionals at the level of the Generalized Gradient Approximation (GGA) and a plane-wave basis set
are widely used today to perform {\em ab initio} molecular dynamics (AIMD) simulations.
Going up in the ladder of accuracy of density functionals from GGA (2$^{\rm nd}$ rung) to hybrid density functionals (4$^{\rm th}$ rung) is much desired pertaining to the accuracy of the latter in describing
structure, dynamics, and energetics of molecular and condensed matter systems.
On the other hand, hybrid density functional based AIMD simulations are about two orders of magnitude slower than GGA based AIMD for  systems containing $\sim$100 atoms using $\sim$100 compute cores.
Two methods, namely MTACE and $s-$MTACE, based on a multiple time step integrator and adaptively compressed exchange operator formalism 
are able to provide a speed-up of about 7$-$9
in performing hybrid density functional based AIMD.
%
%
In this work, we report an implementation of these methods using a task-group based parallelization  within the CPMD program package, with the intention to take advantage of the large number of compute cores available on modern
high-performance computing platforms.
%
%
We present here the boost in performance achieved through this algorithm.
This work also identifies the computational bottleneck in the s-MTACE method, and proposes a way to overcome that.
%
%
%
%
\end{abstract}




\maketitle

\section{Introduction}

{
 Kohn-Sham density functional theory (KS-DFT) and plane wave (PW) based {\it ab initio} molecular dynamics (AIMD) techniques 
 are widely used in investigating structural and dynamical properties of condensed matter systems.\cite{marx-hutter-book,Tuckerman-book,leach,frenkel_smit,md_review_tuckerman} 
The accuracy of the KS-DFT calculations crucially depends on the choice of exchange-correlation (XC) functionals. 
Owing to improved accuracy, hybrid functionals are preferred over the commonly used  XC functionals using the Generalized Gradient Approximation (GGA).\cite{Science_DFT_limitations,PRL_DFT_errors,Chemist's_Guide}
Hybrid functionals incorporate a certain fraction of the Hartree-Fock (HF) exchange to the GGA exchange.\cite{Chemist's_Guide,Martin-book,JCP_B3LYP,JCP_PBE0,JCP_HSE}
They give better prediction of energies, structures, electronic properties, reaction barriers, band gap of solids and dynamical properties of liquids.\cite{JCP_B3LYP,JCP_PBE0,JCP_PBE0_model,JCP_HSE,PCCP_HFX,PCCP_HSE,Galli_RSB_JPCL,Chem_Rev_Cohen,jpcl_2011_hfx,jpcl_2019_hfx,jpcl_2020_hfx,Scuseria:JCP:2008,Adamo:2012,JPCB_AIMD_HFX,JCTC_AIMD_HFX,JCP_AIMD_HFX,Mol_Phy_Car_MLWF,Mol_Phy_Car_MLWF_1,JPCB_water_hfx,sagar_JCC,sagar_JCTC,JCP_sagar}  
%
%
However, the prohibitively high computational cost associated with HF exchange energy evaluation makes the hybrid functionals and PW based AIMD simulations extremely time consuming.\cite{JCP_HFX_Voth} 
This limits the routine use of hybrid functionals and PW based AIMD simulations for large condensed matter systems.
A number of promising strategies have been proposed so far to speed-up such calculations, which can be broadly divided into two categories.
The first set of techniques 
introduce some approximations in the evaluation of the HF exchange energy, thereby reducing the computational cost.
Several works have been reported in this direction 
using localized orbitals,\cite{PRB_Car_Wannier,JCP_AIMD_HFX,Mol_Phy_Car_MLWF,Mol_Phy_Car_MLWF_1,Nature_Car_MLWF,PRL_RSB,JCTC_RSB,JCTC_RSB_1,Galli_RSB_CPL,Galli_RSB_JPCL,Galli_RSB_JPCL1,JCP_sagar,Car_hfx_2019,enabling_part2} 
 multiple time step (MTS) algorithms,\cite{HFX_Hutter_JCP,MTS_AIMD_Ursula,MTS_AIMD_Steele_3,JCP_2019_sagar,sagar_JCC,sagar_JCTC} 
{coordinate scaling,}\cite{JPCL_2018_Bircher,CPC_BIRCHER_2020}
 and other {strategies}.\cite{jctc_Bolnykh_2019,single_precision_hfx,HFX_JB,HFX_Goedecker}
The other group of methods improve the performance 
by employing massively parallel algorithms.\cite{HFX_Curioni,DUCHEMIN_2010,VARINI_2013,BARNES_2017} 
A combination of both the strategies has been also used to achieve remarkable speed-up.\cite{Car_hfx_2019,enabling_part2}

Recently, we proposed an efficient and robust method\cite{JCP_2019_sagar,sagar_JCC} for performing hybrid functionals and PW based AIMD.
We employed a MTS integrator\cite{Tuckerman-book}
scheme based on the adaptively compressed exchange (ACE)~\cite{ACE_Lin,ACE_Lin_1} operator  formalism. 
To take advantage of the ACE operator formalism, we partitioned 
the ionic forces
into computationally cheap fast forces using an approximated ACE operator and computationally costly slow forces due to corrections to the approximated ACE operator. 
We denote this method as MTACE hereafter.
This approach 
provided a significant speed-up in AIMD simulations by decreasing the number of exact exchange evaluations.
Subsequently, we improved the efficiency of this method\cite{sagar_JCTC} %
by employing localized orbitals.
In particular, we used the selected column of the density matrix (SCDM)\cite{SCDM_main} method to obtain localized orbitals, and we used these localized orbitals
to build the ACE operator.\cite{Carnimeo_2019}
We will be denoting this method as s-MTACE.
These methods could speed-up the calculations up to an order of magnitude without 
compromising on accuracy.\cite{sagar_JCTC,sagar_JCC,JCP_2019_sagar}
Both MTACE and s-MTACE  methods are found to reproduce the structure and dynamics of bulk water and free energetics of chemical reactions in solutions correctly.

In the PW based KS-DFT codes, wavefunctions and KS potentials can be in real and/or reciprocal space, and these representations are inter-converted with the help of three dimensional (3D) FFTs.
For the optimal performance of the 3D parallel FFTs, PW implementations use a slab decomposition of the real 3D FFT grids to distribute the data. 
For instance, the 3D FFT grids are distributed along the $X$-direction and the $YZ$ planes are distributed among the MPI tasks (or compute-cores).
Weak scaling performance of such implementations is limited by the number of grid points along the $X$ direction.
For typical DFT calculations today, 
the number of grid points along any direction is few hundreds, 
while the number of available compute cores on any modern day supercomputing resource is of the order of few thousands to millions.
Thus, slab decomposition based FFTs cannot take the full advantage of the large computational 
resources available today.
To overcome this, a task group based parallelization strategy was proposed.\cite{HFX_Curioni,CPG_curioni}
This strategy implemented in the CPMD\cite{cpmd} program is called {\cpgroup}.
In the {\cpgroup} implementation of HF exchange computation, the available processors are divided into several task groups and the array that holds the wavefunctions is replicated among these groups.
The total workload of the HF exchange energy computation is divided into several parts and they are distributed evenly among these task groups.
Finally, a global summation across these groups provides the total contribution to the HF exchange energy.

In the present work, we report the implementations of the MTACE and s-MTACE methods together with {\cpgroup}
within the CPMD program and we present their performance.
In particular, we present the scaling behaviour of the MTACE and s-MTACE methods on a large number of CPU compute cores. %
We will be demonstrating here that a significant improvement in the performance of these methods can be achieved through such an approach. 

}

\section{Methods}

\subsection{HF Exchange Operator}
\label{s:methods:reweight}

\begin{figure*}
	\centering
		\includegraphics[scale=0.5]{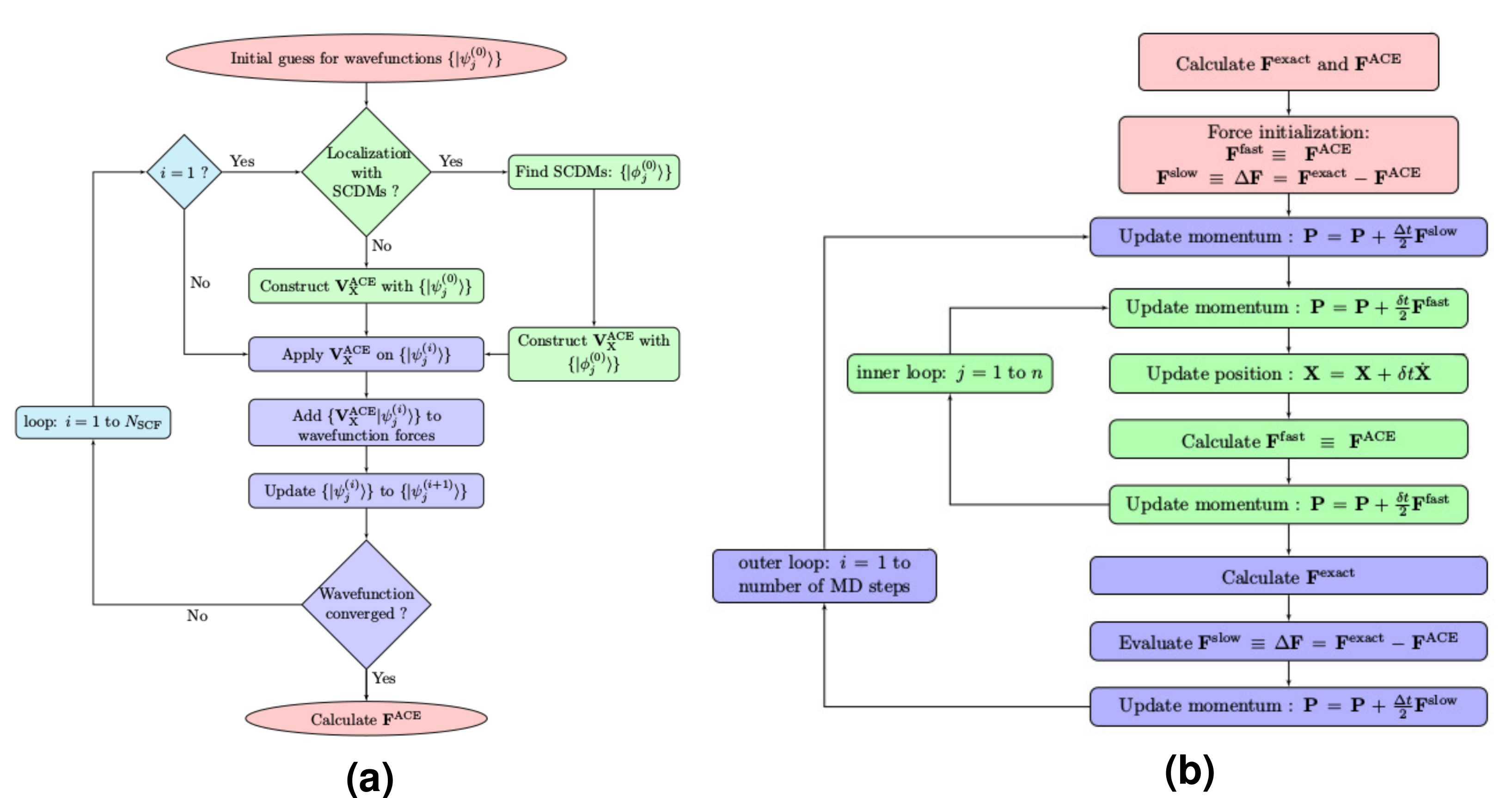}
	\caption{Flowcharts of the MTACE and s-MTACE algorithms: 
	(a) 
	SCF procedure in the s-MTACE method, construction of ${\mathbf V}_{\rm X}^{\rm ACE}$ in the first SCF step is done using SCDM-localized orbitals.
	(b) MTS integration used to perform AIMD simulations.
	}
	\label{FIG:1}
\end{figure*}

In conventional KS-DFT calculations using hybrid density functionals and plane waves, 
evaluation of the HF exchange contributes the most to the total computational time.  
The HF exchange operator ${\mathbf V}_{\rm X}$ is defined as 
\begin{equation}
{\mathbf V}_{\rm X}= -\sum_{j}^{N_{\rm orb}} \frac{| \psi _{j} \rangle  \langle \psi _{j} |}{r_{12}} \enspace ,
\end{equation}
where, $\{|\psi_{j} \rangle\}$ is the set of occupied KS orbitals.
$N_{\rm orb}$ is the total number of occupied orbitals and $r_{12}=\left | \mathbf r_1 - \mathbf r_2 \right | $.
The ${\mathbf V}_{\rm X}$ operator is applied on a KS orbital $|\psi _{i} \rangle$ as
\begin{equation}
\label{vxonpsi}
\begin{split}
{\mathbf V}_{\rm X}|\psi _{i}\rangle & =- \sum_{j}^{N_{\rm orb}} |\psi _{j} \rangle \left \langle\psi _{j} \left | \left ( r_{12}\right )^{-1} \right | \psi _{i}\right \rangle 
\\ & =- \sum_{j}^{N_{\rm orb}} v_{ij}(\mathbf{r}_{1}) |\psi _{j}\rangle \enspace, ~~i=1,....,N_{\rm orb}
\end{split}
\end{equation}
with
\begin{equation}  
\label{e1:vij}
v_{ij}(\mathbf {r}_{1})=\left \langle\psi _{j} \left | \left ( r_{12}\right )^{-1} \right | \psi _{i}\right \rangle \enspace.   
\end{equation}
The HF exchange energy is calculated as
\begin{equation}
\label{e:xc2}
E^{\textrm{HF}}_{\textrm{X}} = - \sum_{i,j}^{N_{\rm orb}}
\left \langle \psi _{i} \left | v_{ij}({\mathbf {r}}_{1}) \right | \psi _{j}
\right \rangle \enspace .
\end{equation}
For an optimal performance, $v_{ij}(\mathbf{r})$ is usually evaluated in reciprocal space.\cite{JCP_HFX_Voth,PRB_Car_Wannier} 
The computational cost for doing Fourier transform scales as $N_{\rm PW}\log  N_{\rm PW}$ using the Fast Fourier Transform (FFT) technique, where $N_{\rm PW}$ is total number of PWs.
Therefore, the total computational cost for the evaluation of exchange energy scales as $N_{\rm orb}^2 N_{\rm PW}\log  N_{\rm PW}$,\cite{JCP_HFX_Voth} as $v_{ij}(\mathbf{r})$ has to be evaluated $N_{\rm orb}^{2}$ times.
This scaling is the reason behind the computational time required for hybrid functional calculations.

\subsection{ACE Operator
}
\label{s:methods:reweight:ts}

Lin~Lin~\cite{ACE_Lin,ACE_Lin_1} introduced 
a low rank decomposition of the  
${\mathbf V}_{\rm X}$ operator 
in the form 
\[ {\mathbf V}_{\rm X}^{\rm ACE}= -\sum_{k}^{N_{\rm orb}}  | P_{k} \rangle  \langle P_{k} | \enspace . \]
Here, the set of ACE projection vectors $\{|P_{k} \rangle\}$ can be computed by a decomposition of the ${\mathbf V}_{\rm X}$ operator, see Ref.\cite{ACE_Lin,JCP_2019_sagar} for details.
The construction of $\{|P_{k} \rangle\}$ requires evaluation of $\{{\mathbf V}_{\rm X}|\psi _{i}\rangle \}$, which follows the same computationally demanding procedure
requiring $N_{\rm orb}^{2}$ evaluations of $v_{ij}(\mathbf{r})$ as discussed
in the previous section.\cite{ACE_Lin} 
However, once the ${\mathbf V}_{\rm X}^{\rm ACE}$ operator is constructed, it  can be easily applied on KS orbitals through the evaluation of $N_{\rm orb}^{2}$ 
{inner} products as
\begin{equation}
{\mathbf V}_{\rm X}^{\rm ACE}|\psi _{i}\rangle=- \sum_{k}^{N_{\rm orb}} |P_{k} \rangle  \left \langle P_k | \psi_{i} \right \rangle , \enspace ~~i=1,....,N_{\rm orb}  \enspace .
\end{equation}
The advantage of the ACE approach lies in the fact that the application of the ${\mathbf V}_{\rm X}^{\rm ACE}$ operator on each KS orbitals consumes much less time as compared to the ${\mathbf V}_{\rm X}$ operator.
Such a low rank decomposition can be used in multiple ways to speed-up
the HF exchange energy calculations.

\subsection{MTACE Method}

The MTACE method introduced by some of the authors of this paper~\cite{JCP_2019_sagar,sagar_JCC} uses the ACE formalism in the framework of the MTS scheme for speeding-up  hybrid functional based AIMD. 
The self consistent field  (SCF) procedure is
modified to take the benefits of the ACE operator.
In the first SCF step, the ACE operator is constructed after the decomposition of $\mathbf V_{\rm X}$,
which is a computationally demanding step as it involves the computation of $\mathbf V_{\rm X}$.
However, for the remaining SCF steps, we use the same ACE operator constructed
at the first SCF step without recalculating it.
After reaching complete SCF convergence, we compute the
ionic forces $\mathbf F^{\rm ACE}$.
It is to be noted that the optimized wavefunction is most certainly different
to the wavefunction which one would obtain if the ACE operator is updated every SCF step. 
As a result, $\mathbf F^{\rm ACE} \neq \mathbf F^{\rm exact}$, where $\mathbf F^{\rm exact}$ is the
ionic force computed using the HF $\mathbf V_{\rm X}$ operator. 
We take care of this difference in forces within the MTS algorithm as explained later.
The flowchart for this modified SCF procedure is shown in Figure~\ref{FIG:1}(a).

In the MTACE scheme,  
%
%
%
 ionic force component can be written 
 as
\begin{equation}
    F^{\rm exact}_K= F^{\textrm{ACE}}_K+ \Delta F_K  \enspace , \enspace K=1,\cdots,3N_{\rm atom}    \enspace 
\end{equation}
with $\Delta F_K = F^{\rm exact}_K -  F^{\textrm{ACE}}_K $.
In our earlier works,\cite{JCP_2019_sagar,sagar_JCC}
we have shown that differences in the ionic force components of $\mathbf F^{\rm exact}$ and $\mathbf F^{\textrm{ACE}}$ are very small and
it is justified to consider
$\mathbf F^{\textrm{ACE}}$ as fast and $\Delta \mathbf F$ as slow: 
\begin{eqnarray}
F^{\rm slow}_K &\equiv& \Delta F_K =  F^{\rm exact}_K -  F^{\textrm{ACE}}_K, ~~\mathrm{and} \nonumber \\[1ex]
F^{\rm fast}_K &\equiv& F^{\rm ACE}_K \enspace .
\end{eqnarray}
Finally, we employ the reversible reference system propagator algorithm (r-RESPA) scheme\cite{r-RESPA} which allows us to compute the computationally costly $\Delta \mathbf F$ less frequently, in fact every $n$ MD steps, as compared to computationally cheaper $\mathbf F^{\textrm{ACE}}$, resulting in speeding-up the calculations as shown in Figure~\ref{FIG:1}(b).
For more details, see Ref.~\cite{JCP_2019_sagar}.
We could achieve a speed-up of about 7 times using this method for a periodic system containing $\sim$100 atoms employing only 120 compute cores.

\subsection{s-MTACE Method}

A modification of the MTACE method, namely s-MTACE, was 
subsequently proposed,\cite{sagar_JCTC} wherein localized SCDM orbitals\cite{SCDM_main} are used for the construction of the $\mathbf V_{\rm X}^{\rm ACE}$ operator, see Figure~\ref{FIG:1}a.
Based on a rank-revealing QR factorization of ${\mathbf \Psi}^{*}$, where ${\mathbf \Psi}$ is
the matrix with all the occupied KS orbitals, the SCDM method constructs linearly
independent columns of the density matrix ${\mathbf \Gamma}={\mathbf \Psi} {\mathbf \Psi}^{*}$
without computing the full  ${\mathbf \Gamma}$ matrix.
The selected columns of ${\mathbf \Gamma}$ are then used to construct the set of orthonormal localized SCDMs
$\{\phi_i\}$.
By screening $\{\phi_i\}$ based on their spatial overlap,  it is possible to achieve a substantial reduction in the number of orbitals
involved in the evaluation of $\{{\mathbf V}_{\rm X}|\psi _{i}\rangle \}$ in Equation~(\ref{vxonpsi}).
We screen the orbitals using the criteria 
\[ \int d\textbf{r} \left | \phi_{i}(\textbf{r}) \phi_{j}^{*}(\textbf{r}) \right | \geqslant \rho_{\textrm {cut}} \enspace .\]
We have reported that this procedure substantially reduces the cost of $\mathbf V_{\rm X}^{\rm ACE}$ operator construction.\cite{sagar_JCTC}
It has been shown that s-MTACE can achieve one order of magnitude speed-up for a system
containing $\sim$100 atoms using 120 compute cores.

 \subsection{s$^\prime$-MTACE Method}

The rank-revealing QR factorization required for the construction of SCDMs is the most time consuming step of the SCDM procedure. 
To speed-up these calculations, we employ parallel ScaLAPACK routines.
However, these procedures scale poorly when using a large number of processors.

Giannozi and co-workers\cite{Carnimeo_2019} proposed a way to 
improve the computational efficiency of 
finding localized SCDMs. 
In this method, one pre-selects a column of the density matrix based on the electron density and the gradient of the electron density.
A column with index $i$ is selected only if
\begin{equation}
    \rho(\mathbf r_i) >  \left \langle \rho \right \rangle \nonumber 
~~\mathrm{and}
~~    \nabla\rho (\mathbf r_i) < \left \langle \nabla\rho \right \rangle \enspace .
\end{equation}
Here, $ \left \langle \rho \right \rangle$ and $\left \langle \nabla\rho \right \rangle$ are the average electron density and average gradient of the electron density over the grid points.
This pre-screening scheme substantially reduces the number of grid points which are to be considered for QR factorization.
Now, the call to the ScaLAPACK routine involves a smaller size matrix $\tilde{\mathbf \Psi}^{*}$, thereby improving the 
performance.
A successful implementation of this procedure is already available in the Quantum ESPRESSO code.\cite{Espresso_hfx,QE_new_JCP} 
We have implemented the same approach in the CPMD program to improve the performance of the
s-MTACE method.
{This 
will be referred as s$^\prime$-MTACE method hereafter.  
}

\subsection{CP Groups Approach}

\begin{figure*}
	\centering
		\includegraphics[scale=.25]{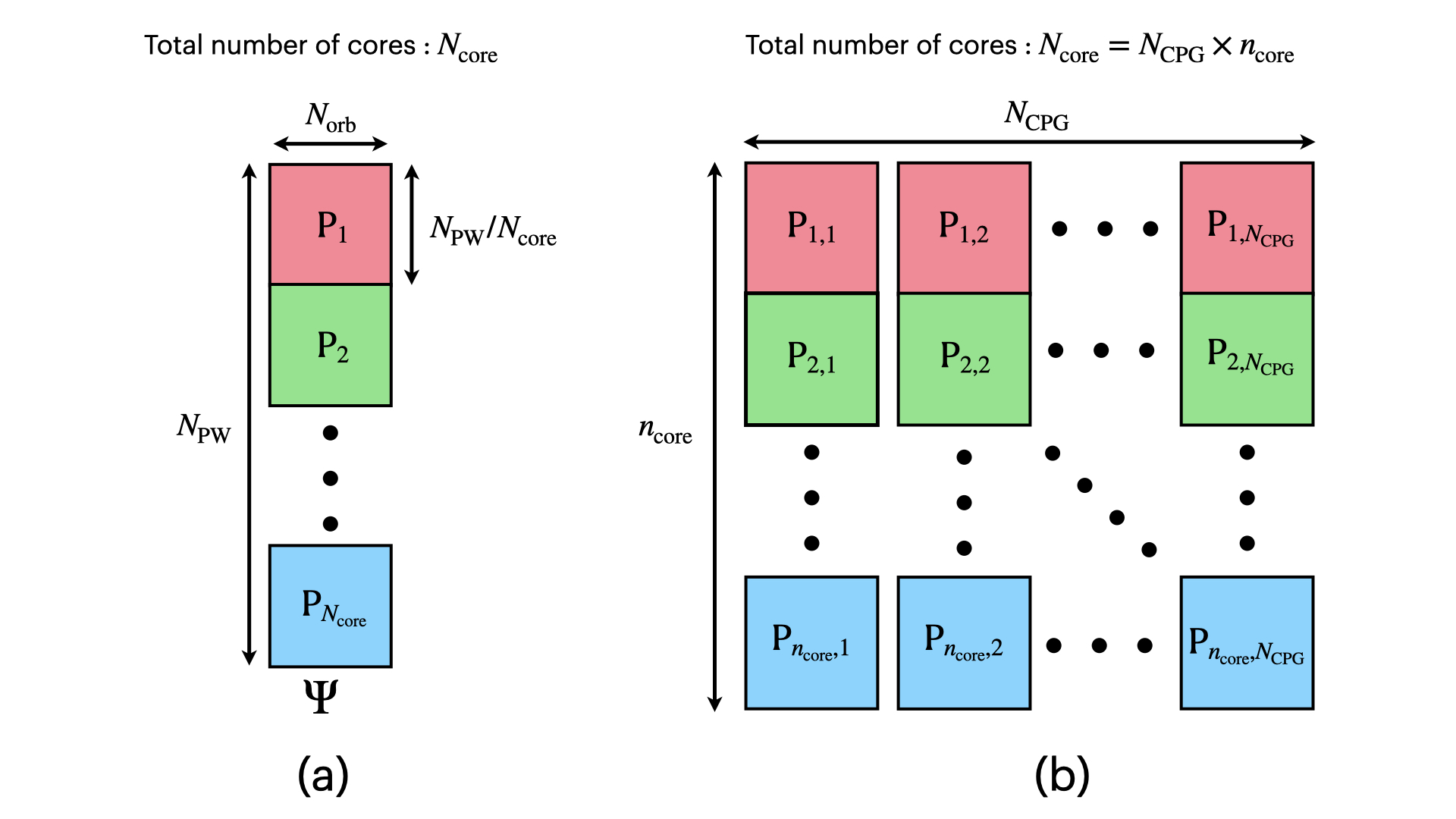}
	\caption{(a) Conventional distribution of the ${\mathbf \Psi}$ matrix with $N_{\rm core}$  compute-cores. The total number of rows ($N_{\rm PW}$) is distributed among these available compute-cores. The part of the matrix residing in different compute-core is shown with different color.
	(b) The {\cpgroup} data distribution is shown for $N_{\rm core}$ compute-cores. Total number of compute-cores are divided into $N_{\rm CPG}$ groups. Each group contains $n_{\rm core}$ compute-cores to distribute the ${\mathbf \Psi}$ matrix. Each compute-core contains $N_{\rm PW}/n_{\rm core}$ rows of the ${\mathbf \Psi}$ matrix.}
	\label{FIG:2}
\end{figure*}

\begin{figure*}
	\centering
		\includegraphics[scale=1.2]{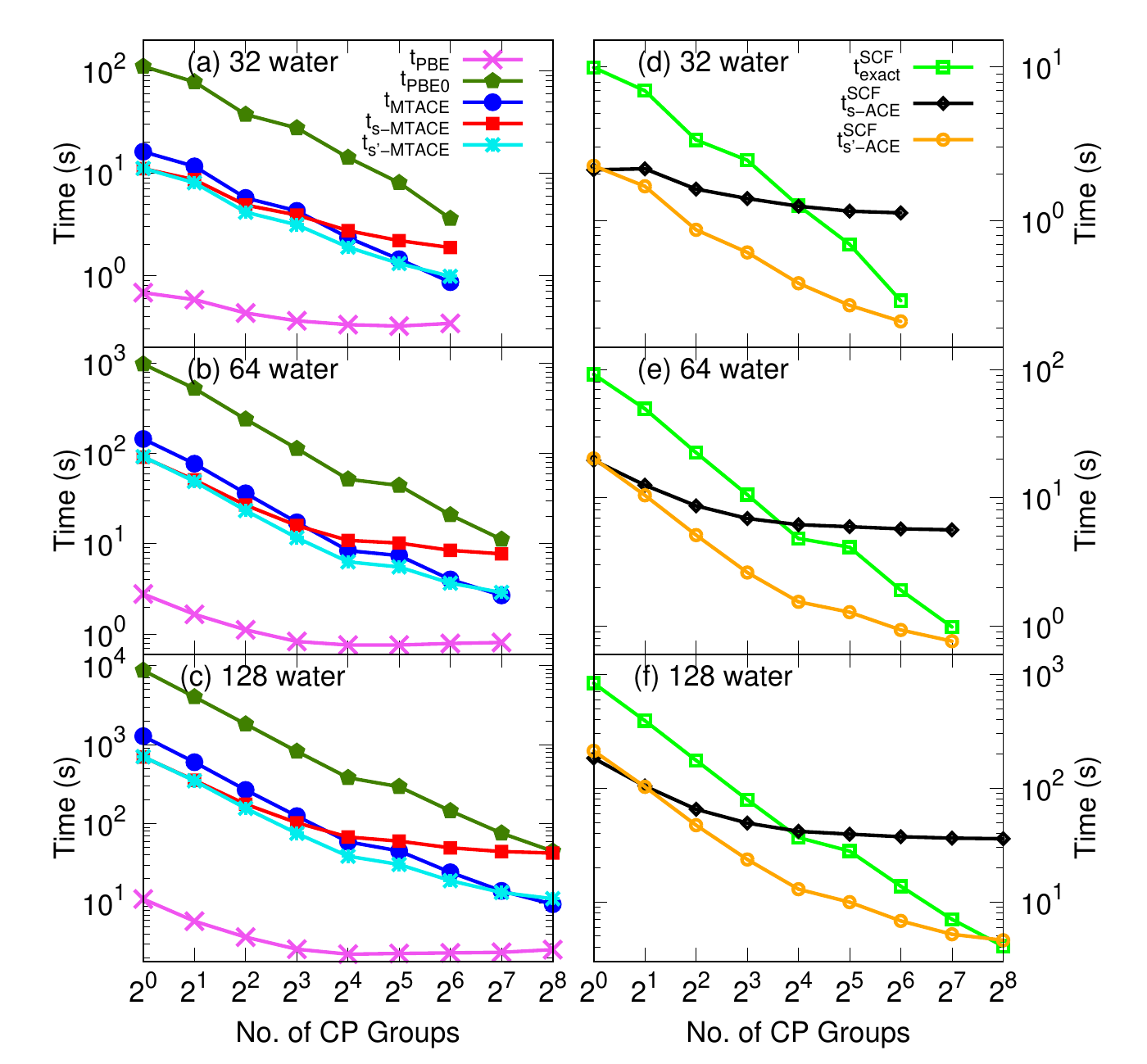}
	\caption{Scaling of the average computational time per BOMD step and average computational time per SCF step for periodic systems containing 32, 64 and 128 water molecules.
$t_{\rm PBE}$, $t_{\rm PBE0}$, $t_{\rm MTACE}$, $t_{\rm s-MTACE}$ and $t_{\rm s'-MTACE}$ are the average computing time per MD step in {\bf PBE}, {\bf PBE0}, {\bf MTACE}, {\bf s-MTACE} and {\bf s'-MTACE} runs.
$t_{\rm exact}^{\rm SCF}$ is the average computing time per SCF step during the computation of $\mathbf F^{\rm exact}$.
$t_{\rm s-ACE}^{\rm SCF}$ and $t_{\rm s'-ACE}^{\rm SCF}$ are the average computing time for the first SCF step during the computation of $\mathbf F^{\rm s-ACE}$ and $\mathbf F^{\rm s'-ACE}$. 
The number of compute cores per {\cpgroup} ($n_{\rm core}$) are 120, 144 and 192 for systems containing 32, 64 and 128 water molecules, respectively.
}
	\label{FIG:3}
\end{figure*}

\begin{figure}
	\centering
		\includegraphics[scale=0.22]{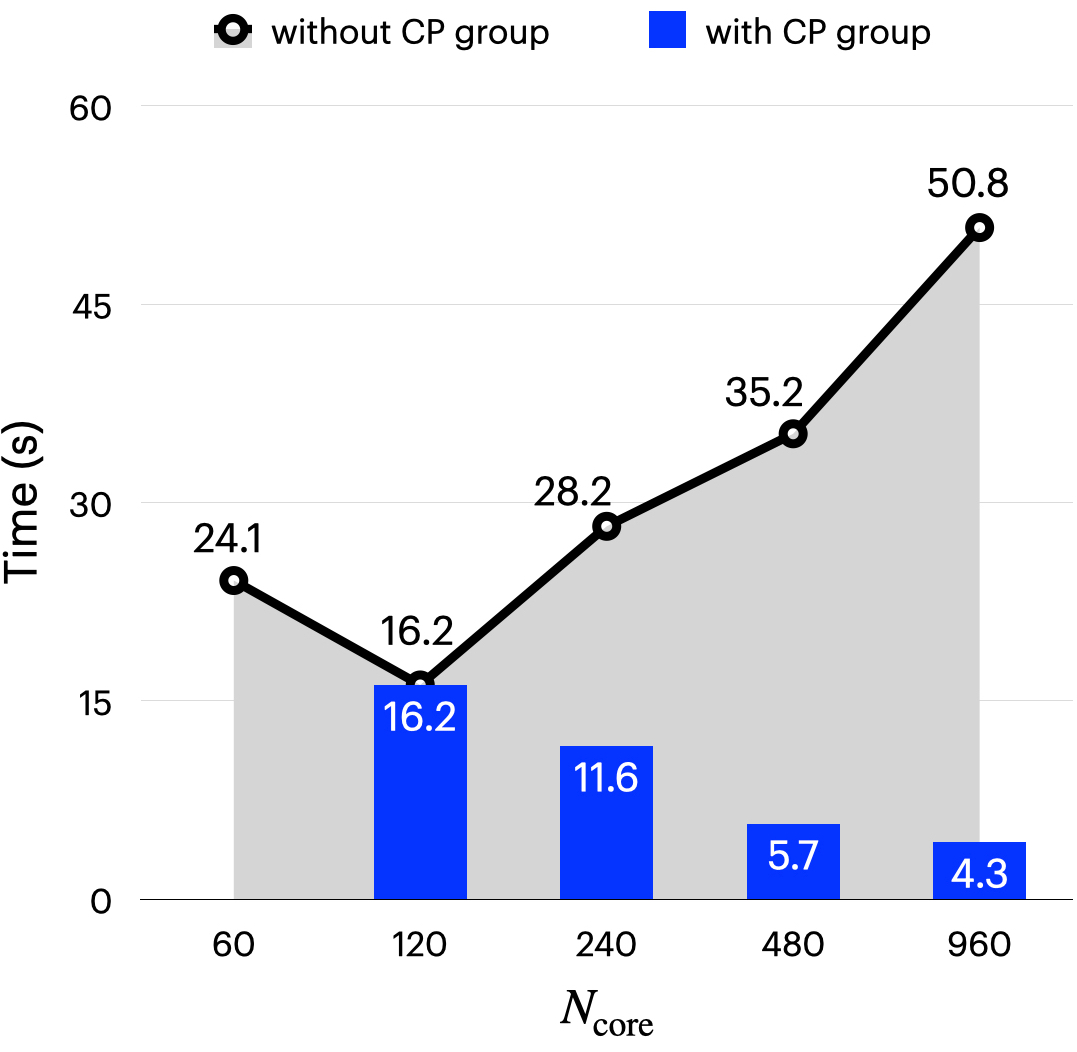}
	\caption{CPU time per MD step for {\bf MTACE} with increasing number of compute cores ($N_{\rm core}$) for a periodic 32-water model.
	Black line shows the scalability of the method without \cpgroup, and the blue bars display the scaling when {\cpgroup} was used. One MD step is 0.48~fs.
	}
	\label{no-cpg-scaling}
\end{figure}

Let us consider that each of the $N_{\rm orb}$ KS orbitals possesses $N_{\rm PW}$ PW coefficients and $N_{\rm core}$
compute-cores are available, then 
each compute core stores $N_{\rm PW}/N_{\rm core}$ rows of
the wavefunction matrix ${\mathbf \Psi}$ in a typical implementation of the slab decomposition, see  Figure~\ref{FIG:2}(a).
In the {\cpgroup} approach, as shown in Figure~\ref{FIG:2}(b), the total number of available compute cores $N_{\rm core}$ divided into $N_{\rm CPG}$ groups.
Each of such a {\cpgroup} possesses $n_{\rm core} = N_{\rm core} / N_{\rm CPG}$ compute cores.
A copy of the whole ${\mathbf \Psi}$ matrix is kept with every group,  distributed among the $n_{\rm core}$ compute cores within that group.
As a result, each compute core of a group keeps $N_{\rm PW}/n_{\rm core}$ rows of the ${\mathbf \Psi}$ matrix. 
The workload across the task groups
 is parallelized over the orbital pairs entering
 the exchange integral in such a manner that computations within each {\cpgroup} is restricted to a subset of orbital pairs.
 Through this, computation of $v_{ij}(\mathbf{r})$ (in Equation~\ref{e1:vij}), which has to be performed for all the orbital pairs, can now be done in chunks across the
 {\cpgroup}.
Finally, a global sum across the groups is performed to 
evaluate the full exchange operator.
The replication of the whole ${\mathbf \Psi}$ matrix among the {\cpgroups} minimizes the inter group communication.
It has been shown that the {\cpgroup} approach can 
be made use to achieve excellent scaling performance in 
hybrid functional based calculations on several thousands of compute cores.\cite{HFX_Curioni,CPG_curioni}

\section{Results and Discussion}
We are presenting here the results of scaling
tests of the MTACE, s-MTACE and s$^\prime-$MTACE methods using the {\cpgroups} implementation.
Benchmark calculations were carried out for periodic supercells with 32, 64, and 128 water molecules (Table~\ref{tbl1}).

\begin{table}
\caption{Details of the liquid water systems used in our benchmarking studies. $N_{\rm water}$ is the number of water molecules, $N_{\rm atom}$ is the number of atoms, $N_{\rm orb}$ is the number of orbitals, $N_{\rm grid}$ is the number of grid points, $N_{\rm PW}^{\rm wave}$ is the number of PWs for wavefunction cutoff, and $N_{\rm PW}^{\rm density}$ is the number of PWs for density cutoff.
}\label{tbl1}
\begin{tabular*}{\tblwidth}{@{} CCCCCCC@{} }
\toprule
$N_{\rm water}$ & $N_{\rm atom}$ & $N_{\rm orb}$ & $N_{\rm grid}$ & Cell size ({\AA}) & $N_{\rm PW}^{\rm wave}$ & $N_{\rm PW}^{\rm density}$ \\
\midrule
32 & 96 & 128 & {120} & 9.85 & 39103 & 311563   \\
64 & 192 & 256 & 144 & 12.41 & 77978 & 623469  \\
128 & 384 & 512 & 192 & 15.64 & 156181 &  1247311  \\
\bottomrule
\end{tabular*}
\end{table}

\begin{table}
\caption{Different simulation runs; 
Note: for the case of MTS based BOMD, we have $n>1$, where $n=\Delta t/ \delta t$.
}\label{tbl:labels}
\begin{tabular*}{\tblwidth}{@{} CCCC @{} }
\toprule
Simulation Label   &    Functional     & $n$   & BOMD Scheme \\
\midrule
{\bf PBE}   & GGA/PBE     &         1 &  Conventional  \\
{\bf PBE0}  & Hybrid/PBE0 &         1 &  Conventional      \\
{\bf MTACE} & Hybrid/PBE0 &     15    & MTACE \\
{\bf s-MTACE} & Hybrid/PBE0 &     15    & s-MTACE  \\
{\bf s'-MTACE} & Hybrid/PBE0 &     15    &  s$^\prime$-MTACE \\
\bottomrule
\end{tabular*}
\end{table}

\begin{table}
\caption{
The number of BOMD steps over which the average compute-times were calculated.
$N_{\rm MD}^{\rm X}$ is the total number of MD steps  with method {X}.
$\rho_{\rm cut}$ is the cutoff used for the screening of the SCDMs in {\bf s-MTACE} and {\bf s$^\prime$-MTACE} runs. $n_{\rm core}$ is the number of compute cores per task group.
}\label{tbl_nmd}
\begin{tabular*}{\tblwidth}{@{} CCCCCCCC@{} }
\toprule
$N_{\rm water}$  & $N_{\rm MD}^{\rm PBE}$ & $N_{\rm MD}^{\rm PBE0}$ & $N_{\rm MD}^{\rm MTACE}$ & $N_{\rm MD}^{\rm s-MTACE}$ & $N_{\rm MD}^{\rm s\prime -MTACE}$ & $\rho_{\rm cut}$ & $n_{\rm core}$ \\
\midrule
32  & 500 & 300 & 300 & 300 & 300 &  $2.5\times10^{-2}$ & 120 \\
64  & 300 & 50 & 150 & 150 & 150  &  $1.0\times10^{-2}$ & 144 \\
128 & 300 & 20 & 75 & 75 & 75 &  $2.0\times10^{-3}$  & 192 \\
\bottomrule
\end{tabular*}
\end{table}

\begin{table*}
\caption{Average computational time per BOMD step and average computational time per SCF step for periodic systems containing 32, 64 and 128 water molecules. 
$N_{\rm water}$ is the number of water molecules, $N_{\rm core}$ is the total number of CPU compute cores, $N_{\rm node}$ is the total number of compute nodes, $N_{\rm CPG}$ is the number of TASK groups.
$t_{\rm PBE}$ is the average computing time per MD step using GGA (PBE) functional.
$t_{\rm PBE0}$, $t_{\rm MTACE}$, $t_{\rm s-MTACE}$ and $t_{\rm s'-MTACE}$ are the average computing time per MD step using {\bf PBE0}, {\bf MTACE}, {\bf s-MTACE} and {\bf s$^\prime-$MTACE} methods.
$t_{\rm exact}^{\rm SCF}$ is the average computing time per SCF step during the computation of $\mathbf F^{\rm exact}$.
$t_{\rm s-ACE}^{\rm SCF}$ and $t_{\rm s'-ACE}^{\rm SCF}$ are the average computing time for the first SCF step during the computation of $\mathbf F^{\rm s-ACE}$ and $\mathbf F^{\rm s'-ACE}$. 
All the times reported are in seconds.
Calculations were done with 48 compute cores per node, except for the case with $N_{\rm water}=32$ and $N_{\rm CPG}=1$ case where 24 compute cores per node were used. }
\label{tbl3}
\begin{tabular*}{\tblwidth}{@{} LCCCCCCCCCCC@{} }
\toprule
$N_{\rm water}$ & $N_{\rm core}$ & $N_{\rm node}$ & $N_{\rm CPG}$ & $t_{\rm PBE}$   & $t_{\rm PBE0}$  & $t_{\rm MTACE}$   & $t_{\rm s-MTACE}$   & $t_{\rm s'-MTACE}$   & $t_{\rm exact}^{\rm SCF}$  & $t_{\rm s-ACE}^{\rm SCF}$  & $t_{\rm s'-ACE}^{\rm SCF}$   \\
\midrule
32 & 120 & 5 & 1  & 0.68  & 110.19 & 16.20 & 11.05 & 11.12  & 9.92 & 2.14 & 2.27 \\
 & 240 & 5 & 2   & 0.58 & 77.88 & 11.62 & 8.61 & 8.06  & 7.00 & 2.17 & 1.67 \\
 & 480 & 10 & 4   & 0.43 & 37.38 & 5.71 & 4.88 & 4.15  & 3.35 & 1.60 & 0.87 \\
 & 960 & 20 & 8   & 0.36 & 27.64 & 4.27 & 3.90 & 3.12  & 2.47 & 1.39 & 0.62 \\
 & 1920 & 40 & 16   & 0.33  & 14.19 & 2.34 & 2.74 & 1.89  & 1.25 & 1.24 & 0.39 \\
 & 3840 & 80 & 32   & 0.32  & 8.04 & 1.44 & 2.19 & 1.31  & 0.70 & 1.15 & 0.28 \\
 & 7680 & 160 & 64  & 0.34  & 3.60 & 0.86 & 1.87 & 0.98  & 0.30 & 1.12 & 0.22 \\
\midrule
 64 & 144 & 3 & 1  & 2.79  & 974.08 & 144.42  & 90.77 & 92.14 & 91.98 & 19.61 & 20.27 \\
  & 288 & 6 & 2  & 1.66 & 523.59 & 76.90  & 50.97 & 48.85 & 49.42 & 12.52 & 10.46 \\
  & 576 & 12 & 4  & 1.12  & 239.36 & 36.25  & 26.86 & 23.38 & 22.52 & 8.66 & 5.11 \\
  & 1152 & 24 & 8   & 0.83 & 113.13 & 17.28  & 15.93 & 11.66  & 10.55 & 6.90 & 2.61 \\
  & 2304 & 48 & 16  & 0.76  & 51.79 & 8.40  & 10.92 & 6.31  & 4.81 & 6.16 & 1.54 \\
  & 4608 & 96 & 32  & 0.76 & 44.14 & 7.40 & 10.20 & 5.55  & 4.11 & 5.95 & 1.28 \\
  & 9216 & 192 & 64  & 0.79  & 20.91 & 4.04  & 8.46 & 3.67  & 1.90 & 5.72 & 0.93 \\
  & 18432 & 384 & 128  & 0.81  & 11.19 & 2.69  & 7.73 & 2.90  & 0.98 & 5.62 & 0.76 \\
\midrule
128  & 192 & 4 & 1   & 11.12  & 8776.64 & 1285.86 & 699.56 & 708.00 & 837.51 & 183.01 & 212.59 \\
  & 384 & 8 & 2  & 5.83  & 4039.04 & 600.21 & 355.45 & 348.78 & 391.59 & 104.11 & 102.83 \\
  & 768 & 16 & 4  & 3.66  & 1829.83 & 268.16 & 176.99 & 157.06 & 175.60 & 65.23 & 47.26 \\
  & 1536 & 32 & 8  & 2.57 & 823.68 & 124.82 & 102.89 & 75.66  & 78.96 & 49.24 & 23.61 \\
  & 3072 & 64 & 16  & 2.22 & 383.01 & 58.84 & 67.84 & 38.82  & 36.80 & 41.70 & 12.94 \\
  & 6144 & 128 & 32   & 2.27 & 294.42 & 45.25 & 60.08 & 30.44  & 27.96 & 39.44 & 9.95 \\
  & 12288 & 256 & 64  & 2.31 & 145.67 & 24.25 & 49.56 & 19.02  & 13.73 & 37.41 & 6.81 \\
  & 24576 & 512 & 128  & 2.35 & 75.79 & 14.11 & 44.41 & 13.43  & 7.03 & 36.32 & 5.20 \\
  & 49152 & 1024 & 256   & 2.54  & 44.80 & 9.55 & 42.37 & 11.22  & 4.08 & 35.92 & 4.61 \\
\bottomrule
\end{tabular*}
\end{table*}
\begin{figure}
	\centering
		\includegraphics[scale=0.18]{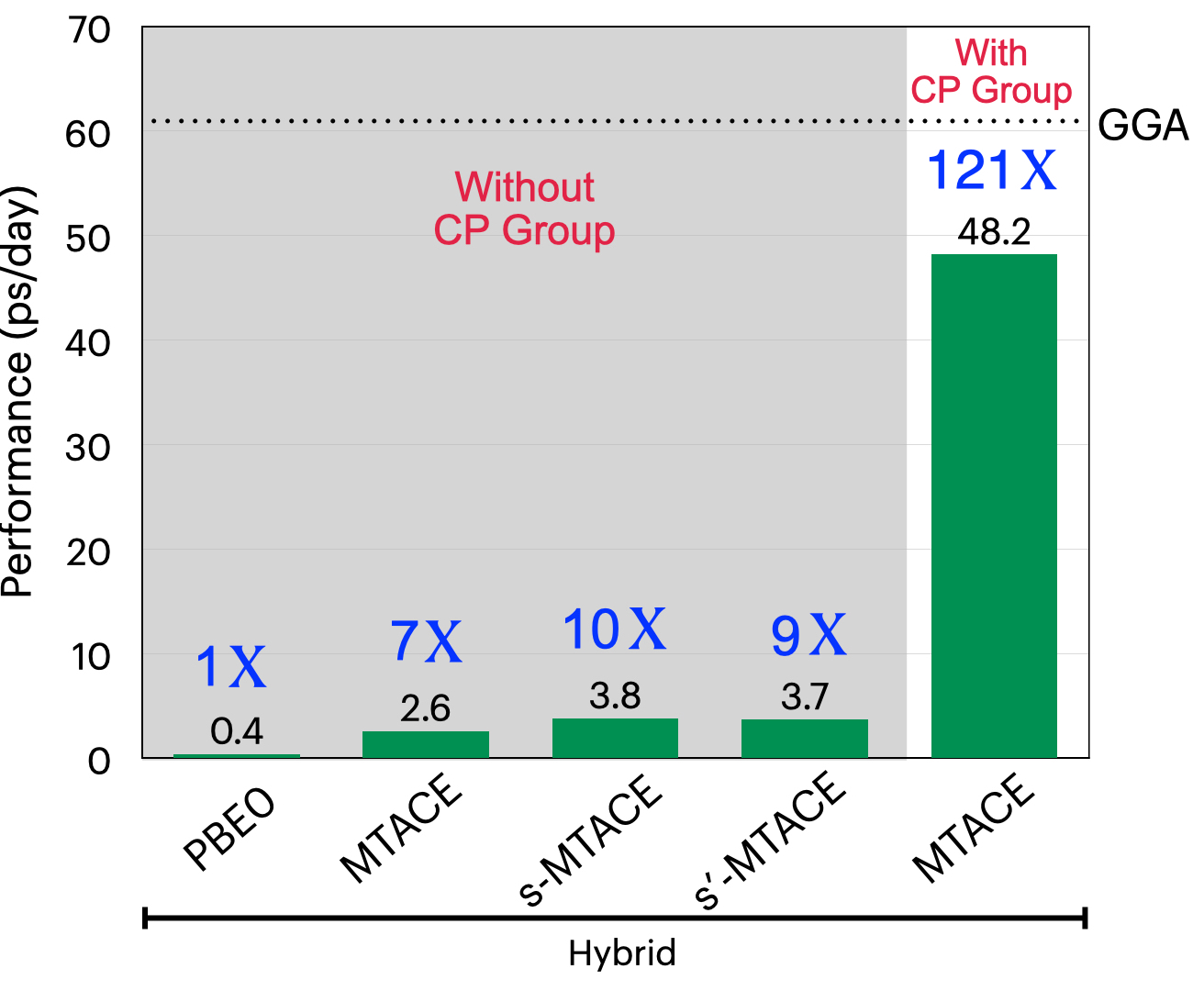}
	\caption{ Best performance of the methods discussed in this article for the 32 water system.
	Performance has been measured as the length of the trajectory (in ps) generated in one day. 
	A time step of 0.48~fs was only used as H atom mass was taken as 1~amu.
	Grey area highlights the schemes where {\cpgroup} was not used. 
	120 compute cores were used for all the calculations. 
	Dotted black line indicates the performance of the GGA calculations with 120 compute cores.
	The best performance with {\cpgroup}  was obtained for the {\bf MTACE} on 7680 compute cores.
	Effective speed-up compared to PBE0 is also indicated here (blue).
	}
	\label{best-perf}
\end{figure}

\begin{table*}[]
\caption{Decomposition of total computational time per SCF step for periodic systems containing 32, 64 and 128 water molecules. 
Various contributions to $t_{\rm s-ACE}^{\rm SCF}$ and $t_{\rm s'-ACE}^{\rm SCF}$ are reported. 
$t_{\rm SCDM}$ or $t_{\rm SCDM}^{'}$ is the time for the localization procedure.
%
$t_{\rm QR}$ is the time for the QR factorization and $t_{\rm other}$ is the compute time for the rest of the part.
$t_{\rm comput}$ is the compute time for the actual computation of the HF exchange energy.
All compute times reported here are in seconds.
}\label{tbl4}
\begin{tabular}{cc|cccc|cccc}
\toprule
\multicolumn{1}{c}{\multirow{2}{*}{$N_{\rm water}$}} & \multicolumn{1}{c|}{\multirow{2}{*}{$N_{\rm CPG}$}} & \multicolumn{1}{c}{\multirow{2}{*}{$t_{\rm s-ACE}^{\rm SCF}$}} & \multicolumn{2}{c}{$t_{\rm SCDM}$}                          & \multicolumn{1}{c|}{\multirow{2}{*}{$t_{\rm comput}$}} & \multicolumn{1}{c}{\multirow{2}{*}{$t_{\rm s'-ACE}^{\rm SCF}$}} & \multicolumn{2}{c}{$t_{\rm SCDM}^{'}$}                          & \multicolumn{1}{c}{\multirow{2}{*}{$t_{\rm comput}$}} \\ \cline{4-5} \cline{8-9}
\multicolumn{1}{c}{}                   & \multicolumn{1}{c|}{}                   & \multicolumn{1}{c}{}                   & \multicolumn{1}{c}{$t_{\rm QR}$} & \multicolumn{1}{c}{$t_{\rm other}$} & \multicolumn{1}{c|}{}                   & \multicolumn{1}{c}{}                   & \multicolumn{1}{c}{$t_{\rm QR}$} & \multicolumn{1}{c}{$t_{\rm other}$} & \multicolumn{1}{c}{}                   \\  \midrule
\multirow{7}{*}{32}     &   1    &  2.14  &   0.46   &  0.04  & 1.65 &  2.27 &  0.00  & 0.03 & 2.23  \\
 &      2    &  2.17 &    0.93   &  0.06  & 1.18  &  1.67   &  0.01  &  0.05   & 1.62  \\
&  4  & 1.60 &  0.94  & 0.06  & 0.60  &    0.87 &  0.01 &  0.05  &  0.82  \\
  & 8   &   1.39   &   0.93  & 0.06 & 0.40  &   0.62   & 0.01  & 0.05  & 0.56 \\
    &   16 &    1.24     &  0.94  & 0.05   & 0.25    &   0.39  & 0.01 & 0.05 &  0.33  \\
   &    32 & 1.15    &   0.93  & 0.05  & 0.17  &   0.28   & 0.01 & 0.05  &  0.22 \\
   &     64     &  1.12    &   0.94   &  0.05  & 0.13  & 0.22 & 0.01  &  0.05 & 0.16 \\  \midrule 
\multirow{8}{*}{64}      &    1   &  19.61   & 5.10 &  0.24  &  14.27  &  20.27   & 0.01  &  0.20 &   20.06    \\
        &   2    &  12.52   &  5.00  & 0.23 & 7.29  &  10.46   & 0.01  & 0.19 & 10.25 \\
        &   4    &  8.66  &  4.92   &   0.21  & 3.52 &  5.11 & 0.02  &  0.19  & 4.90 \\
        &    8   &  6.90   &  4.90   & 0.22 & 1.78  & 2.61 & 0.02  & 0.19    &  2.40 \\
        &    16   & 6.16 &  4.89  & 0.22 & 1.05 &  1.54  &  0.02 &  0.19  & 1.33 \\
        &    32   & 5.95 &  4.89  & 0.21 & 0.85  & 1.28  &  0.03 &  0.18  & 1.07 \\
        &     64  & 5.72 &  4.90  & 0.21 & 0.61 & 0.93  &  0.04 &  0.18 &  0.70  \\
        &   128    & 5.62 &  4.90  & 0.21  & 0.51  &  0.76 & 0.04  & 0.18 &  0.53    \\  \midrule 
\multirow{9}{*}{128}        &   1    &  183.01  &  32.57 & 1.26  & 149.18  &  212.59   & 0.13 & 1.03  & 211.43   \\
        &   2    &  104.11   & 32.01 & 1.24 &  70.87 & 102.83  & 0.13 &  1.02  & 101.68   \\
        &    4   &  65.23 &  31.72  &  1.19  & 32.32 & 47.26 &  0.13 &  1.01  & 46.11 \\
        &    8   &   49.24    &  31.67   & 1.19 & 16.38  & 23.61 &  0.13 &  1.01  & 22.46  \\
        &    16   & 41.70 &  31.64 & 1.18 & 8.89 & 12.94 &  0.14 &  1.01  & 11.79  \\
        &    32   & 39.44 & 31.63 & 1.17 & 6.65  & 9.95 &  0.15 &  0.99 &  8.81 \\
        &     64  &  37.41   &  31.64  & 1.17 & 4.60 & 6.81 &  0.16   &  0.99  & 5.66   \\
        &   128    & 36.32 & 31.65  &  1.17  & 3.50 & 5.20  &  0.18  &  0.98  & 4.03  \\
        &    256   & 35.92 &  31.69  & 1.17 & 3.06  & 4.61  & 0.26  &   0.99 &  3.36  \\ \bottomrule 
        
\end{tabular}
\end{table*}

\begin{figure*}
	\centering
		\includegraphics[scale=0.25]{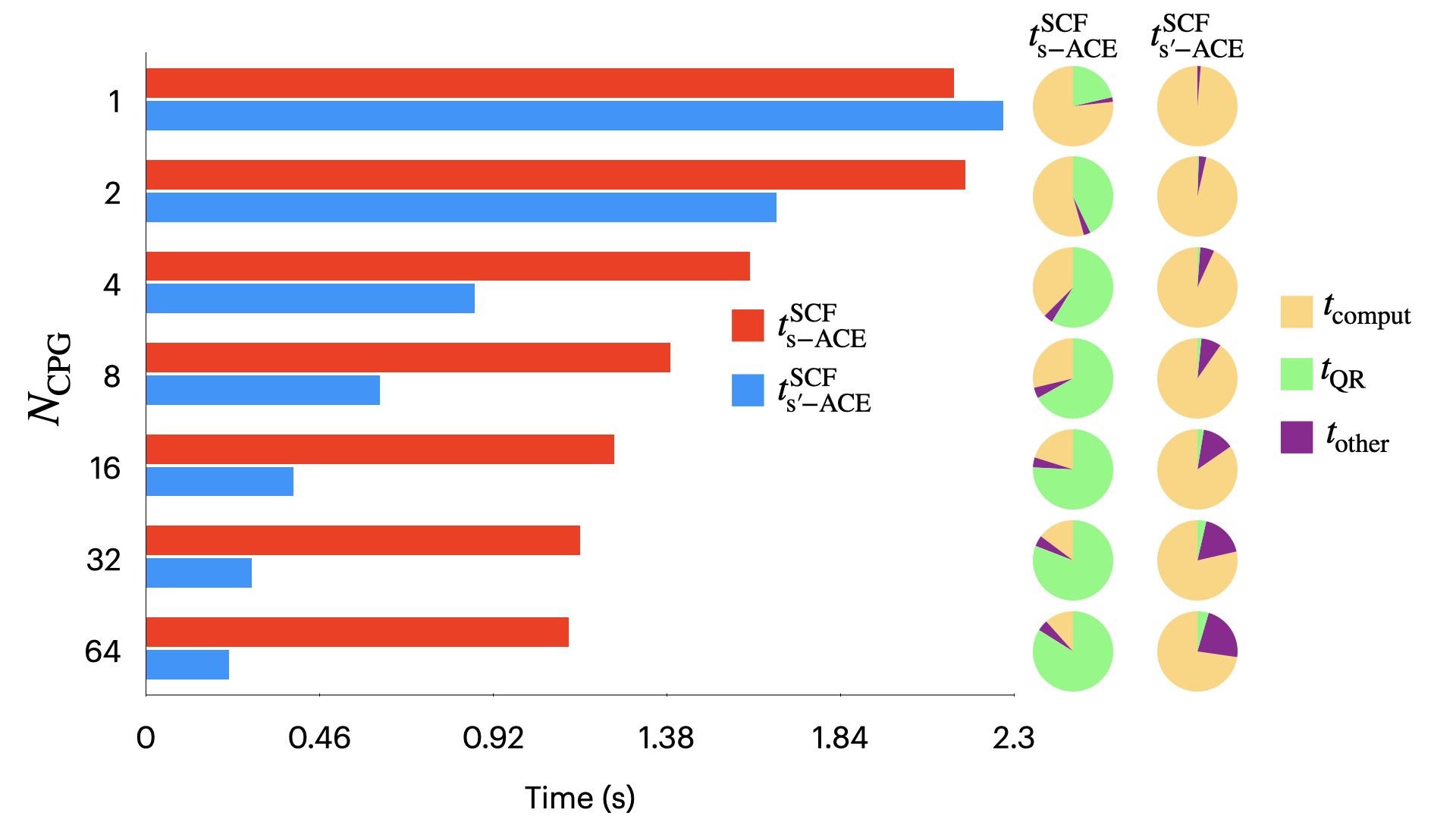}
	\caption{Scaling performance of computational time per SCF step for {\bf s-MTACE} and {\bf s$^\prime$-MTACE} with number of {\cpgroup} ($N_{\rm CPG}$) for the periodic 32 water model.
	Pie charts show the percentage of time spent for different contributions in one SCF step.
    Here, $t_{\rm QR}$ is the computational time for QR factorization, $t_{\rm other}$ is the computational time for the rest,
    %
    and $t_{\rm comput}$ is the time for the actual computation of the HF exchange energy.
	}
	\label{SCF-decompose}
\end{figure*}

\begin{figure*}
	\centering
		\includegraphics[scale=0.25]{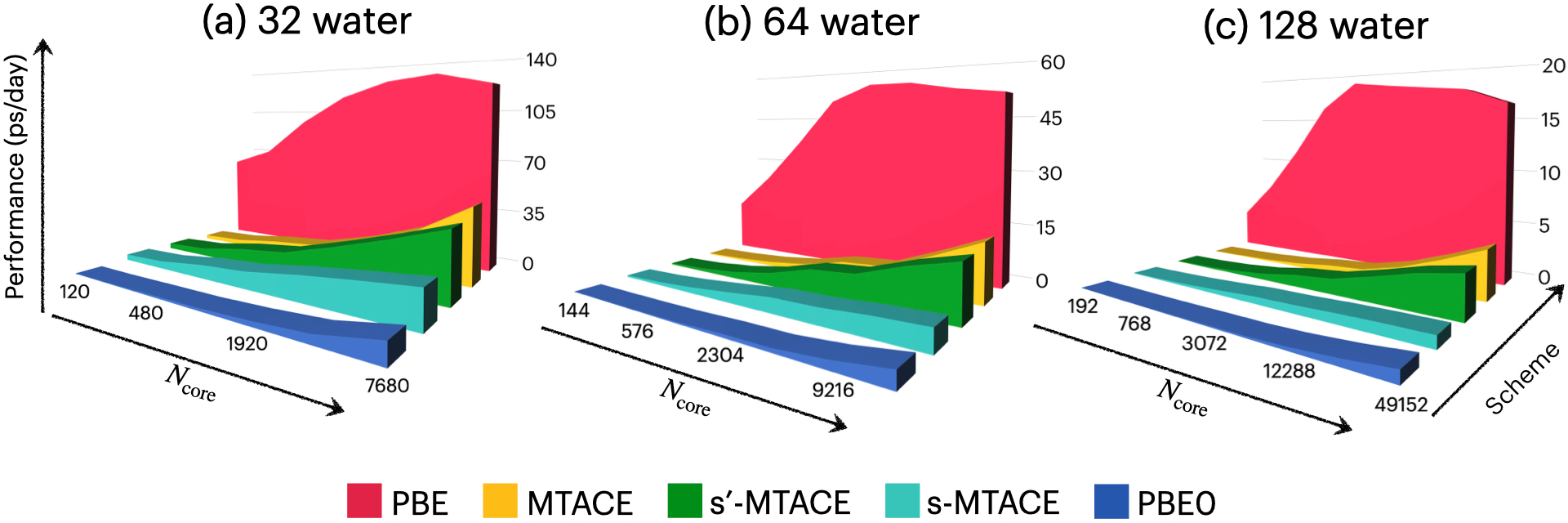}
	\caption{Performance measured in units of ps per day for periodic systems containing (a) 32, (b) 64 and (c) 128 water molecules.
	%
	$X$ axis denotes number of compute cores ($N_{\rm core}$) and
	$Y$ axis is indicating different methods described in this article.
	$Z$ axis is ps of trajectory that can be generated per day using all of these schemes.
	}
	\label{perf}
\end{figure*}
All the methods and algorithms presented earlier were implemented in { a modified version of}
the {\tt CPMD 4.3} program,\cite{cpmd,KLOFFEL2021} 
and adapted for the existing {\cpgroup} implementation within the program.
We used PBE0~\cite{JCP_PBE0_model} (hybrid) and PBE~\cite{PRL_GGA_PBE} (GGA) functionals 
for all the computations.
Core electrons were accounted by using 
norm-conserving Troullier Martin pseudopotentials.~\cite{PRB_TM}
A cutoff energy of 80~Ry was used to expand the wavefunctions in the PW basis set.
We carried out Born-Oppenheimer molecular dynamics (BOMD) simulations and the wavefunction convergence criteria in SCF calculations was set to $10^{-6}$ au for the wavefunction gradients.
At every MD step, initial guess of the wavefunctions was constructed based on the Always Stable Predictor Corrector extrapolation scheme\cite{JCC_ASPC} with order 5.
For standard PBE and PBE0 calculations, the standard Velocity Verlet scheme was employed with a timestep of $\Delta t = 0.48$~fs.
However, for the PBE0 runs with the MTS scheme,  $\delta t = 0.48$~fs and  $\Delta t = 7.2$~fs (i.e., $n=15$) were taken as the smaller and larger time steps, respectively.
We use the labels {\bf PBE}, {\bf PBE0}, {\bf MTACE},  {\bf s-MTACE}, and  {\bf s$^\prime$-MTACE} 
to indicate different methods used in this work, see Table~\ref{tbl:labels}.

All the benchmark calculations presented here were performed on SuperMUC-NG located at Leibniz Supercomputing Center (LRZ).  
The compute nodes are equipped with two Intel\textsuperscript{\textregistered} Skylake Xeon Platinum 8174 processors (24 compute cores per processor). 
Each compute node has 48 compute cores and 96~GB memory.
The nodes are interconnected through a fast Intel\textsuperscript{\textregistered} OmniPath network with 100 {\tt Gbit/s} speed.

In Table~\ref{tbl_nmd}, we give the number of steps for which average compute timings were calculated
and the values of $\rho_{\rm cut}$ for the s-MTACE runs. 
For an ideal load balancing, the number of compute cores per {\cpgroup} ($n_{\rm core}$) is chosen to be equal to the number of grid points in the $X$-direction.
We chose $n_{\rm core}$ as 120, 144 and 192 for systems containing 32, 64 and 128 water molecules, respectively, see Table~\ref{tbl_nmd}.
The average computational time per BOMD step is reported in  Table~\ref{tbl3} for the {\bf PBE}, {\bf PBE0}, {\bf MTACE} and {\bf s-MTACE} methods.
Also, the scaling is shown in Figure~\ref{FIG:3}.
First, we observe that the {\bf PBE} calculations have poor scaling with {\cpgroups} for all the systems.
This is expected as these calculations lack enough computationally scalable work that can be distributed over the {\cpgroups}.
In fact due to the extra overhead of communication and synchronisation, we notice small increase in 
computational time per MD step with large number of {\cpgroups}.
However, for all the other three methods which use hybrid functionals, we notice  considerable improvements in the performance with increase in {\cpgroups}.
Noticeably, {\bf PBE0}  scales almost perfectly with the number of {\cpgroups}. 
The scaling behavior of {\bf MTACE} is also 
as good as {\bf PBE0}.
It is clear that the scaling of {\bf MTACE} without the {\cpgroup} approach is poor when more than 120 compute cores 
are used for the 32 water system (Figure~\ref{no-cpg-scaling}).

%
From Table~\ref{tbl3}, we observe that the ratio of the average computing time per MD step for {\bf MTACE} and {\bf PBE} ($t_{\rm MTACE}/t_{\rm GGA}$) decreases with increasing number of \cpgroup.
This is a consequence of the fact that the {\bf MTACE} calculations are scaling well with {\cpgroups} as compared to {\bf PBE}.
Depending on the system size, {\bf MTACE} is only 2-4 times slower than {\bf PBE} runs by employing the combination of MTACE and {\cpgroups} when a sufficiently large number of processors
is used.
The best case scenarios for the systems with 32, 64 and 128 water molecules are having
{\bf MTACE} runs clocking only 2.7, 3.5 and 4.3 times slower than {\bf PBE}, respectively.
Further, {\bf MTACE} is giving a 4-5 fold speed-up compared to {\bf PBE0} when using the highest number of {\cpgroups} we employed.
For a 32 water system, the computing time for one MD step is now only 0.86~s for hybrid
functional based BOMD with the {MTACE} method when 7680 compute-cores were taken.
Figure~\ref{best-perf} shows the improved performance of the method with and without the {\cpgroup} approach.
Similar enhancement in performance is also seen for
systems with 64 and 128 water molecules.
These results are encouraging as we can generate long trajectories at the level of hybrid functionals within a shorter time by making use of large computing resources.
The best computing performances we obtained are 48, 15 and 4 ps of trajectory per day for systems with 32, 64 and 128 water molecules, respectively (see also Figure~\ref{perf}).
We would like to emphasis that $\delta t=0.48$~fs was taken considering that H atoms were assigned 1~amu mass.
By using a deuterium mass for H atoms, a $\delta t=1$~fs could be used, resulting in doubling the simulation performance (ps/day).

Our calculations show that the scaling behavior of {\bf s-MTACE} deteriorates beyond a certain number of compute cores.
In order to understand this poor scaling behavior, we have looked at the average computing time per SCF step during different modes of force calculations.
We label the average computing times per SCF step during the computation of $\mathbf F^{\rm exact}$
and $\mathbf F^{\rm s-ACE}$ as $t_{\rm exact}^{\rm SCF}$, and 
$t_{\rm s-ACE}^{\rm SCF}$, respectively.
We observe that $t_{\rm s-ACE}^{\rm SCF}$ scales poorly when {\cpgroups} is large (Table ~\ref{tbl3}). 
To scrutinize the poor scaling of $t_{\rm s-ACE}^{\rm SCF}$, we decomposed the time for various stages of computation (Table ~\ref{tbl4} and Figure~\ref{SCF-decompose}).
The parallel QR factorization with the ScaLAPACK routines ($t_{\rm QR}$) contributes mostly to the
computational overhead for the SCDM localization procedure ($t_{\rm SCDM}$).
It is clear that $t_{\rm QR}$ scales poorly with $N_{\rm CPG}$, resulting in an overall poor scaling of $t_{\rm SCDM}$.
On the other hand, computation of the HF exchange integrals ($t_{\rm comput}$) scales 
well with  $N_{\rm CPG}$.
%
%

%
To overcome this and to improve the scalability
we adopted the {s$^\prime$-MTACE} method.
%
The reported results in Table~\ref{tbl4} and Figure~\ref{SCF-decompose} suggest that the QR factorization with the ScaLAPACK routine now consumes negligible amount of computing time.
As a result,the  poor scaling of $t_{\rm QR}$ has no significant effect on
the overall scaling of the method.
It has to be noted that the pre-screening procedure used in the {s$^\prime$-MTACE} 
method slightly deteriorates the localization properties of the computed SCDMs, resulting 
in a large number of overlapping pairs during the construction of the ACE operator.
Consequently, $t_{\rm comput}$ turns out to be higher than that of the s-MTACE scheme.
The overall scaling behavior of MD timings ($t_{\rm s^\prime-MTACE}$) is satisfactory as
can be seen in Table~\ref{tbl3} and Figure~\ref{FIG:3}.

\section{Summary}
%
We have presented a detailed benchmarking study on the computational performance of the
{\bf MTACE} and {\bf s-MTACE} algorithms with task-groups (\cpgroup) for performing AIMD simulations 
with hybrid density functionals and plane waves.
In our implementations of the {\bf MTACE} and {\bf s-MTACE} methods using the {\cpgroup} environment in the CPMD program,  orbital pairs are distributed across the processor groups to achieve a better scaling performance.
Through this implementation, we are able to accomplish excellent scaling behavior beyond $\sim$100 compute cores, 
even for typical system sizes with $\sim$100 atoms. 
Further, excellent speed-up has  been also seen while using this implementation.
In the best performance achieved for a model system containing 32 water molecules,  
computational overhead for doing hybrid density functional based AIMD is only 3 times more expensive than with 
GGA. 
%
Our implementation has resulted in boosting
the performance of hybrid functional based AIMD of
this system by a factor of 121 (see Figure~\ref{best-perf}).
%
%
The performance of the {s-MTACE} method was better than the {MTACE} for a small number of {\cpgroups}, however, it deteriorated with increasing number of the {\cpgroups}.
This problem was overcome by the implementation of the { s$^\prime$-MTACE} method.
%
Our results suggest that either the {MTACE} or {s$^\prime$-MTACE} method in combination with {\cpgroups}
is ideal for running hybrid density functional based AIMD simulations on high-performance computers.
For system with finite band gap, {s$^\prime$-MTACE} should perform better than {MTACE}.
%

\section*{Acknowledgement}
{Financial support from the National Supercomputing Mission (Subgroup Materials and Computational Chemistry) and Science and Engineering Research Board (India)
under the MATRICS scheme (Ref. No. MTR/2019/000359) and from the German
Research Foundation (DFG) through Research Unit FOR 1878 (funCOS) and
Collaborative Research Center SFB 953 (project number 182849149) are
gratefully acknowledged.
RK thanks the Council of Scientific \& Industrial Research (CSIR), India for
her Junior Research Fellowship (JRF).
Computational resources were provided by 
SuperMUC-NG (project pn98fa) at Leibniz Supercomputing Centre (LRZ).
}


\bibliographystyle{apsrev} 

\bibliography{cas-refs}

\end{sloppypar} 
\end{document}